\documentclass[11pt,reqno]{amsproc}
\usepackage[colorinlistoftodos,shadow]{todonotes}
\usepackage[T1]{fontenc}
\usepackage{amsmath,amscd,amssymb}
\usepackage{cite}
\usepackage{color}
\usepackage{graphicx}
\usepackage{psfrag,epsfig}
\usepackage{multirow}
\usepackage{rotating}
\usepackage[letterpaper,margin=0.75in]{geometry}
\usepackage{subfigure}
\usepackage{enumerate}
\usepackage{comment}
\usepackage{lineno}
\usepackage{algorithmic}
\usepackage{algorithm}
\usepackage[small]{caption}
\usepackage{bbold}
\usepackage[debug=false, colorlinks=true, pdfstartview=FitV, linkcolor=blue, citecolor=blue, urlcolor=blue]{hyperref}
\numberwithin{equation}{section}
\usepackage{chemarr}
\usepackage[version=3]{mhchem}
\newcommand*\patchAmsMathEnvironmentForLineno[1]{%
  \expandafter\let\csname old#1\expandafter\endcsname\csname #1\endcsname
  \expandafter\let\csname oldend#1\expandafter\endcsname\csname end#1\endcsname
  \renewenvironment{#1}%
     {\linenomath\csname old#1\endcsname}%
     {\csname oldend#1\endcsname\endlinenomath}}%
\newcommand*\patchBothAmsMathEnvironmentsForLineno[1]{%
  \patchAmsMathEnvironmentForLineno{#1}%
  \patchAmsMathEnvironmentForLineno{#1*}}%
\AtBeginDocument{%
\patchBothAmsMathEnvironmentsForLineno{equation}%
\patchBothAmsMathEnvironmentsForLineno{align}%
\patchBothAmsMathEnvironmentsForLineno{flalign}%
\patchBothAmsMathEnvironmentsForLineno{alignat}%
\patchBothAmsMathEnvironmentsForLineno{gather}%
\patchBothAmsMathEnvironmentsForLineno{multline}%
}

\title[PFLOTRAN-SIP for geoelectrical signatures]{PFLOTRAN-SIP: A~PFLOTRAN Module for Simulating Spectral-induced Polarization of Electrical Impedance Data}
\author[B.~Ahmmed et~al.]{B.~Ahmmed$^{1,2}$, M.\,K.~Mudunuru$^2$, S.~Karra$^2$, S.\,C.~James$^{3,*}$, H.\,S.~Viswanathan$^2$, and J.\,A.~Dunbar$^1$ \\
{\scriptsize $^{1}$Department of Geosciences, Baylor University, Waco, TX 76706.} \\
{\scriptsize $^{2}$Earth and Environmental Sciences Division, Los Alamos National Laboratory, Los Alamos, NM 87545.} \\
{\scriptsize $^{3}$Departments of Geosciences and Mechanical Engineering, Baylor University, Waco, TX 76706.} \\
}
\thanks{$^*$Corresponding author: \texttt{sc\_james@baylor.edu}.}
\date{\today}
\begin{document}
\maketitle
\let\thefootnote\relax\footnotetext{\textbf{Authorship statement:} B. Ahmmed developed the framework, ran models, and drafted the original manuscript. M.\,K. Mudunuru formulated the idea, supervised, and helped draft the manuscript. S. Karra wrote the code and analyzed data. S.\,C. James supervised, participated in drafting, and critically revised the manuscript. H.\,S. Viswanathan critically revised the manuscript. J.\,A. Dunbar supervised and analyzed data.}
\section*{ABSTRACT}
\label{Sec:abstract}
Spectral induced polarization (SIP) is a non-intrusive geophysical method that gleans information in the form of chargeability (the ability of a material to retain charge) in the time domain or its phase shift in the frequency domain.
SIP is widely used to detect sulfide minerals, clay minerals, metallic objects, municipal wastes, hydrocarbons, and salinity intrusion.
Although SIP is a temporal method, it cannot measure the dynamics of flow and solute/species transport in the subsurface over long times (often 10--100s of years).
To capture long-term dynamics, data collected with the SIP technique need to be coupled with fluid flow and reactive-transport models.
To our knowledge, there is no simulator in the open-source literature that couples fluid flow, solute transport, and SIP process models to analyze geoelectrical signatures in large-scale systems.
\texttt{PFLOTRAN-SIP} was built to couple SIP data to fluid flow and solute transport processes. 
This framework built on the \texttt{PFLOTRAN}-\texttt{E4D} simulator that couples \texttt{PFLOTRAN} (a massively parallel multi-physics simulator for subsurface flow and transport) and \texttt{E4D} (a massively parallel geoelectrical simulator) without sacrificing computational performance.
\texttt{PFLOTRAN} solves the coupled flow and solute transport process models to estimate solute concentrations, which were used in Archie's model to compute bulk electrical conductivities at near-zero frequency.
These bulk electrical conductivities were modified using the Cole-Cole model to account for frequency dependence.
Using the estimated frequency-dependent bulk conductivities, \texttt{E4D} simulated the real and complex electrical potential signals for selected frequencies for SIP.
These frequency-dependent bulk conductivities contain information relevant to geochemical changes in the system.
The \texttt{PFLOTRAN-SIP} framework was demonstrated through a synthetic tracer-transport model simulating tracer concentrations and electrical impedance for four frequencies.
Later, SIP inversion estimated bulk electrical conductivities by matching electrical impedance for each of these frequencies.
Estimated bulk electrical conductivities were consistent with the simulated tracer concentrations from the \texttt{PFLOTRAN-SIP} forward model.
This framework allows practitioners of environmental hydrogeophysics and biogeophysics to monitor chemical, nuclear, and tracer transport sites as well as to detect sulfide minerals, metallic objects, municipal wastes, hydrocarbons, and salinity intrusion.
\\
\\
\textbf{Keywords:}~Electrical resistivity tomography, inversion
spectral induced polarization (SIP),
subsurface geoelectrical signatures,
hydrogeophysics,
multi-physics,
porous media flow.

%

\section{INTRODUCTION}
\label{Sec:S1_Intro}
Engineered subsurface systems are dynamic due to natural and anthropogenic activities that alter porosity, permeability, fluid saturation, and geochemical properties over time \cite{national2000research}. 
Geophysical techniques such as seismic (deep or near-surface seismic) and potential-based methods (electromagnetic, magnetic, electrical resistivity tomography [ERT], spectral induced polarization [SIP]) characterize changes in the subsurface \cite{Revil2011,kearey2013,snieder2007advanced}.
Among these, ERT and SIP map the distributions of bulk electrical conductivity (i.e.,~the reciprocal of resistivity) due to changes in subsurface fluid flow, temperature, deformation, and reactive transport \cite{carpenter2017stress,kaselow2004stress,gresse2017,Robinson2015,Byrdina2013}. 
Because structural, topological, and geochemical properties (e.g.,~pore structures, fracture networks, electron donor, etc.) influence bulk electrical conductivity \cite{Revil2011,snieder2007advanced}, ERT and SIP are applied in environmental and energy industries to characterize subsurface interactions. 
Hence, coupling ERT and/or SIP process models to flow and reactive-transport process models can enhance the interrogation of engineered subsurface systems.

ERT's data-collection component measures the electric potentials resulting from an applied direct current (DC) while the data-processing component inverts these measured potentials to map the spatial distribution of bulk electrical conductivities \cite{Byrdina2013,Revil2011,Revil2004}.
ERT looks at amplitude responses not their frequencies; therefore, it is difficult to capture multi-frequency data (typically, greater than 20~Hz).
Because subsurface properties are often frequency-dependent, ERT fails to interrogate the polarization features of geologic materials, heavy metals, and induced-polarization minerals (e.g., clay minerals, hydrothermal-alteration products, pyrite, finely disseminated sulfide minerals, etc.) \cite{yanetal2014, he2005hydrocarbon,Revil2011}. 
However, by injecting alternating currents (AC), the induced polarization (IP) method can measure ``chargeability'' in the time domain or ``phase shift'' in the frequency domain, which is the phase angle (phase lag) between the applied current and induced voltage of polarized geologic materials \cite{sumner2012principles,vaulet2011}.
The IP method measures the energy storage capacity of certain minerals and can be used to detect hydrocarbons \cite{luo1988}, contaminant plumes \cite{olhoeft1986direct,morgan1999induced,vanhala1997mapping}, municipal waste, green waste (agricultural and biodegradable wastes) \cite{landfill}, sulfide minerals \cite{yanetal2014,butler2005}, and hydrothermal products \cite{yanetal2014,butler2005}.
IP is a single- or double-frequency method that generally fails to distinguish between a true IP response (e.g.,~polarized geologic materials) and noise (e.g.,~electromagnetic interference) \cite{butler2005,luo1988}.
IP signals are often low in porous geologic media and noise often masks polarization responses.
Moreover, polarization responses are frequency-dependent and reach their maxima at different frequencies.
Therefore, SIP data collected at multiple frequencies improve subsurface imaging even under noisy conditions.

SIP is representative of a polarization response and cannot directly measure contaminant concentrations or chemical reactions.
Coupling with a subsurface flow and reactive-transport model can tie SIP back to these processes.
Furthermore, SIP is a temporal method to image subsurface bulk electrical conductivities. 
However, in practice, subsurface contaminant transport is a slow process ($\approx1-100$~years).
A~continuous SIP survey across a wide range of frequencies is feasible over such a long time.
As a result, SIP is performed at discrete times (snapshots) and for discrete frequencies.
The acquired data are then tied back to subsurface processes through coupling to flow and reactive-transport models.
The electrical conductivity from the SIP method contains information on the spatial distribution of conducting fluids and fluid chemistry.
In addition, the SIP method inverts for frequency-dependent electrical conductivity based on measured/simulated electrical-impedance and phase-shift data, which facilitates detection, extraction, and understanding of the evolution of hydrogeophysical and biogeophysical signatures at both lab and field scales \cite{KEMNA2012,Atekwana2009,mellage2018linking,SLATER2002}.

While there are numerous software to model geoelectrical data (e.g., \texttt{Res2Dinv} \cite{loke1996rapid,loke2003comparison,res3dinv}, \texttt{Aarhusinv} \cite{Aarhusinv2013}, \texttt{BERT} \cite{rucker2006three, gunther2006three}, \texttt{EarthImager3D} \cite{agi2008}, \texttt{E4D} \cite{johnsonetal2010}, \texttt{pyGIMLi}  \cite{Ruecker2017}, and \texttt{ZondRes3D} \cite{zond3res}), none capture the physics associated with dynamic subsurface processes.
These software packages can also image frequency-dependent electrical conductivities but cannot capture dynamic subsurface processes.
To overcome these problems, Johnson et~al. \cite{JOHNSON2017} developed the massively parallel \texttt{PFLOTRAN-E4D} simulator, which couples \texttt{PFLOTRAN} \cite{Hammondetal2014}, a subsurface flow and reactive-transport code, to \texttt{E4D}, a finite element code for simulating and inverting geoelectrical data.
However, \texttt{PFLOTRAN-E4D} does not account for induced polarization.
To capture dynamics of subsurface processes and the true sources of induced polarization, a computationally efficient framework is needed to couple fluid flow and solute transport with the SIP process model.
\emph{This work extended the capabilities of \texttt{PFLOTRAN-E4D} to include SIP in a framework called \texttt{PFLOTRAN-SIP}.}
Here, in a medium with polarization properties, \texttt{PFLOTRAN-SIP} was demonstrated with a representative tracer-transport model.

\section{PFLOTRAN-SIP:~PROCESS MODELS AND COUPLING FRAMEWORK}
\label{Sec:S2_coupling}
The \texttt{PFLOTRAN-SIP} framework couples flow and reactive-transport process models in \texttt{PFLOTRAN} \cite{hammond2012pflotran,Hammondetal2014,pflotran-web-page,pflotran-user-ref} with the SIP process model in \texttt{E4D} \cite{e4dinweb,johnsonetal2010,e4dmanual} to characterize fluid-driven electrical impedance signatures across multiple frequencies.
At each time-step, simulation outputs from \texttt{PFLOTRAN} (fluid saturation, tracer concentration, etc.) were supplied to Archie's Law \cite{archie1942} to calculate fluid-dependent bulk electrical conductivities for \texttt{E4D} simulations.
These estimated bulk electrical conductivities were decomposed into real and imaginary components for each frequency using the Cole-Cole model \cite{colecole1941,tarasov2013use}, which is an empirical description of frequency-dependent behavior of bulk electrical conductivities.
These processes were repeated until the entire transient simulation was completed.

\subsection{E4D Process Model}
\label{SubSec:S2_E4D_SIP_Model}
\texttt{E4D} is an open-source, massively parallel, finite-element code for simulating and inverting three-dimensional time-lapsed ERT and SIP data \cite{e4dinweb,e4dmanual,johnsonetal2010,johnson20173}.
The process models in \texttt{E4D} for ERT and SIP assume that displacement currents are negligible and current density can be described by Ohm's constitutive model \cite{johnson20173}.
These assumptions result in a Poisson equation relating induced current to the electric potential field:
\begin{equation}
  \label{eq:e4d_poissons}
  -\mathrm{div} \left[\sigma \left(\mathbf{x} \right) \mathrm{grad}
  \left[\Phi_\mathrm{\sigma}\left(\mathbf{x}\right)\right] \right]=
  \mathcal{I} \delta \left(\mathbf{x} - \mathbf{x}_0 \right),
\end{equation}
where $\mathrm{\sigma}$ [S/m] is the effective electrical conductivity, $\mathcal{I}$ [A] is the injected current, and $\Phi_\sigma(\mathbf{x})$\,[V] the electrical potential all at position-vector $\mathbf{x}$\,[m], $\delta$\,($\cdot$) is the Dirac delta function, div is the divergence of a vector field while grad is the gradient of a scalar field \cite{lipnikov2014mimetic,droniou2014finite}.

Eq.~(\ref{eq:e4d_poissons}) models the DC effect, which is required in ERT forward/inverse modeling; however, it does not account for induced polarization under AC.
IP under AC results in a secondary potential that needs to be accounted for in the SIP forward/inverse modeling.
This requires modification of Eq.~(\ref{eq:e4d_poissons}) to solve for the total electrical potential field under IP effects:
\begin{equation}
  \label{eq:ip_potential_field}
  -\mathrm{div} \left[ \left(1 - \eta\left(\mathbf{x}\right) \right)
  \sigma \left(\mathbf{x} \right) \mathrm{grad}
  \left[\Phi_\eta\left(\mathbf{x}\right)\right] \right] =
  \mathcal{I} \delta \left(\mathbf{x} - \mathbf{x}_0 \right),
\end{equation}
where $\Phi_\eta$\,[$\mathrm{V}$] is the total electrical potential field, which includes IP effects from a polarized material with chargeability distribution $\mathbf{\eta\left(r\right)}$\,[milliradians] \cite{seigel1959}.
The secondary potential resulting from the IP effect is \cite{oldenburg1994}:
\begin{equation}
  \label{eq:secondary_potential}
  \Phi_\mathrm{s} = \Phi_\eta - \Phi_\sigma,
\end{equation}
and the apparent chargeability is \cite{seigel1959}:
\begin{equation}
  \label{eq:apparent_chargeability}
  \eta_\mathrm{a} = \frac{\Phi_\eta - \Phi_\sigma}{\Phi_\eta}.
\end{equation}
Secondary potential $\Phi_\mathrm{s}$ and apparent chargeability $\eta_\mathrm{a}$ are weakly nonlinear that result from Eqs.~(\ref{eq:e4d_poissons}) and (\ref{eq:ip_potential_field}).
These potentials $\Phi_\eta$, $\Phi_\sigma$, and $\Phi_\mathrm{s}$ are time-domain signatures of induced polarization.
Eq.~(\ref{eq:secondary_potential}) is in the time domain and is transformed into the frequency domain:
\begin{equation}
  \label{eq:ip_potential_field_freq}
  -\mathrm{div} \left[ \sigma^*\left(\mathbf{x},\mathrm{\omega}\right) \mathrm{grad}
  \left[\Phi^*\left(\mathbf{x}\right)\right] \right] =
  \mathcal{I} \delta \left(\mathbf{x} - \mathbf{x}_0 \right),
\end{equation}
where $\omega$\,[Hz] is the frequency.
$\sigma^*(\mathbf{x},\mathrm{\omega})$\,[S/m] and $\Phi^*(\mathbf{x})$\,[V] are the frequency-dependent electrical conductivities and electrical potential, respectively.
$\Phi^*(\mathbf{x})$ consists of real and imaginary electrical potentials corresponding to induced polarization.
Zero potential is enforced on boundaries of the domain \cite[Section~3]{johnson20173} to solve Eq.~(\ref{eq:ip_potential_field_freq}).

\texttt{E4D} simulates four-electrode configurations (e.g., Wenner and dipole-dipole arrays) \cite{johnsonetal2010}.
Current is injected from source to sink electrodes while measurements are recorded between the other two electrodes \cite{johnsonetal2010,kearey2013,Robinson2015}.
For ERT, the measured response is the potential difference (voltage) between the two electrodes while SIP also includes the phase shift (radians).
Based on the user-defined survey design, \texttt{E4D} can simulate up to thousands of ERT/SIP measurements to compute electrical potential distributions.
Because the governing equations are linear in $\Phi_{\sigma}$ and $\Phi_\eta$, \texttt{E4D} solves Eq.~(\ref{eq:ip_potential_field_freq}) by superimposing pole solutions with different current sources that makes ERT or SIP forward modeling highly scalable \cite{johnsonetal2010,johnson20173}.

\texttt{E4D} solves the ERT and SIP process models in the frequency domain using a low-order finite element method (FEM).
The output of the FEM solution for the ERT process model is electrical potential throughout the domain, which is real valued and frequency independent.
Because the SIP process model is frequency dependent, the corresponding output of the FEM solution has both real and imaginary components of electrical potential.
The complex-valued electrical potential (or equivalently the phase-shift distribution in the model domain) provides new information on IP in the subsurface, which is not capturable by ERT.

\texttt{E4D} uses the standard Galerkin weak formulation \cite{hughes2012finite} on an unstructured, low-order, tetrahedral, finite element mesh \cite{tetgen2015} and it iteratively computes the total electrical potential field due to IP effects \cite[Section~3]{johnson20173}.
Equations for computing the real and imaginary components of the complex-valued electrical potential are decoupled, and the finite-element analysis is performed in the real-number domain.
First, \texttt{E4D} solves for the real component without considering IP effects.
Second, the current source for the imaginary component is computed from the real component.
Third, the imaginary component of the total electrical potential is calculated based on this computed current source.
Fourth, the secondary current source arising from the imaginary component is computed.
This secondary current source considers IP effects.
Later, the real component is calculated based on this secondary current source.
These steps are repeated until a convergence criterion is satisfied.

\subsection{PFLOTRAN Process Models}
\label{SubSec:S2_PFLOTRAN_Model}
\texttt{PFLOTRAN} solves a system of nonlinear partial differential equations describing multiphase, multicomponent, reactive flow and transport using the finite-volume method (FVM) \cite{hammond2012pflotran,Hammondetal2014,lichtner2015pflotran}.
In this paper, we consider only single-phase fluid flow and solute transport when predicting the spatio-temporal distribution of solute concentrations.
Mass conservation for single-phase, variably saturated flow is:
\begin{equation}
  \label{eq:Richards_eqn}
  \frac{\partial \phi s \rho}{\partial t} + \mathrm{div}\left[ \rho \mathbf{q} \right] = Q_\mathrm{w},
\end{equation}
where $\rho$\,[kg/m\textsuperscript{3}] is the fluid density, $\phi$\,[--] is the porosity, $s$\,[--] is the saturation, t\,[$\mathrm{s}$] is time, $\mathbf{q}$\,[m/s] is the Darcy flux, and $Q_\mathrm{w}$ [kg/m\textsuperscript{3}/s] is the volumetric source/sink term.
Darcy flux is:
\begin{equation}
  \label{eq:Darcy_velocity}
  \mathbf{q} = - \frac{\kappa \kappa_\mathrm{r}\left(s\right)}{\mu} \mathrm{grad}\left[p - \rho gz \right],
\end{equation}
where $\kappa$\,[m\textsuperscript{2}] is the intrinsic permeability, $\kappa_\mathrm{r}$\,[--] is the relative permeability, $\mu$\,[Pa $\cdot$ s] is dynamic viscosity, $p$\,[Pa] is pressure, $g$\, [m/s\textsuperscript{2}] is gravity, and $z$\,[m] is the vertical component of $\mathbf{x}$.
The source/sink term is:
\begin{align}
  \label{Eqn:SourceSink_Richards_Eqn}
  Q_\mathrm{w} = \frac{q_\mathrm{M}}{W_\mathrm{w}} \delta \left(\mathbf{x} -
  \mathbf{x}_Q \right),
\end{align}
where $q_\mathrm{M}$\,[kg/m\textsuperscript{3}/s] is the mass flow rate, $W_\mathrm{w}$\,[kg/kmol] is the formula weight of water, and $\mathbf{x}_Q$\,[m] denotes the location of the source/sink.
The governing equation for tracer transport is:
\begin{equation}
  \label{eq:advective_dispersion}
  \frac{\partial \phi c}{\partial t}  + \mathrm{div} \left[c\mathbf{q} - \phi s \tau D \, \mathrm{grad}\left[c\right] \right] = Q_c,
\end{equation}
where $c$\,[molality] is the solute concentration, $D$\,[m\textsuperscript{2}/s] is the diffusion/dispersion coefficient, $\tau$\,$[-]$ is tortuosity (path length of the fluid flow), and $Q_c$\,[molality/s] is the solute source/sink term.
Dirichlet, Neumann, or Robin boundary conditions are specified when solving Eqs.~(\ref{eq:Richards_eqn})--(\ref{eq:advective_dispersion}).

\begin{figure}[htbp]
  \centering
  \includegraphics[width=0.5\textwidth]{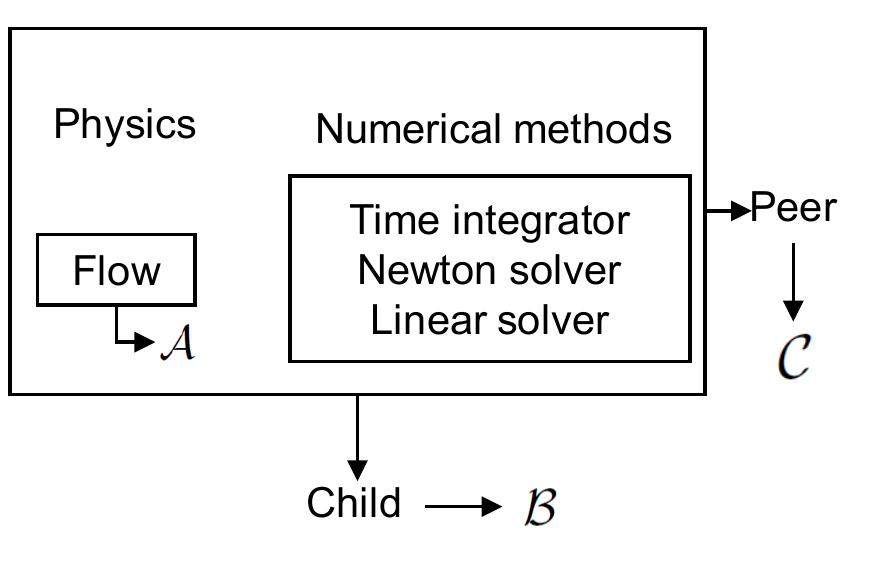}
  \caption{\textbf{\textrm{PFLOTRAN Process Modeling:}} Peer and child process model class of \texttt{PFLOTRAN} (redrawn from \cite{JOHNSON2017}).}
  \label{fig:child_peer_process}
\end{figure}

Coupled governing Eqs.~(\ref{eq:Richards_eqn})--(\ref{eq:advective_dispersion}) are solved with a two-point flux FVM in space and a fully implicit backward Euler method in time using a Newton–Krylov solver \cite{hammond2012pflotran,balay2017petsc}.
\texttt{PFLOTRAN}'s process model tree shown in Fig.~\ref{fig:child_peer_process} has master process $\mathcal{A}$ and pointers to child process $\mathcal{B}$ and peer process $\mathcal{C}$.
Here, the flow model is master process $\mathcal{A}$ while $\mathcal{B}$ and $\mathcal{C}$ are the solute transport and E4D/SIP models, respectively.
The time step for the flow model may be different from solute-transport model.
Transfer of information between $\mathcal{A}$ (e.g.,~flow) and $\mathcal{B}$ (e.g.,~solute transport) takes place before and after each of $\mathcal{A}$'s time steps.
Synchronization of $\mathcal{A}$ and $\mathcal{C}$ (e.g.,~ERT or SIP) occurs at specified times.
Execution starts with the master-process model $\mathcal{A}$, which can take as many adaptive time steps as needed to reach the synchronization point.
$\mathcal{B}$ and $\mathcal{C}$ proceed according to their time steps ($\leq \mathcal{A}$'s) to reach the synchronization point.
When $\mathcal{A}$, $\mathcal{B}$, and $\mathcal{C}$ all reach the synchronization point, variables and parameters (e.g.,~saturation, solute concentration, porosity, etc.) are updated between $\mathcal{A}$ and $\mathcal{C}$.

\subsection{PFLOTRAN-SIP Coupling}
\label{SubSec:S2_PFLOTRAN_SIP_Coupling}
Coupling involves six steps:
(1)~\texttt{PFLOTRAN}'s flow model calculates fluid pressure, saturation, and velocity; (2)~using those simulated outputs, the transport model calculates solute concentrations; (3)~solute concentrations in each \texttt{PFLOTRAN} mesh cell are used to calculate DC electrical conductivities for ERT based on Archie's law; (4)~the Cole-Cole model is used to calculate frequency-dependent electrical conductivities; (5)~real and imaginary electrical conductivities are interpolated onto the \texttt{E4D} mesh; and (6)~the SIP model solves the forward problem to calculate electrical impedance and phase shifts.

\begin{figure}[htbp]
  \centering
  \includegraphics[width=0.5\textwidth]{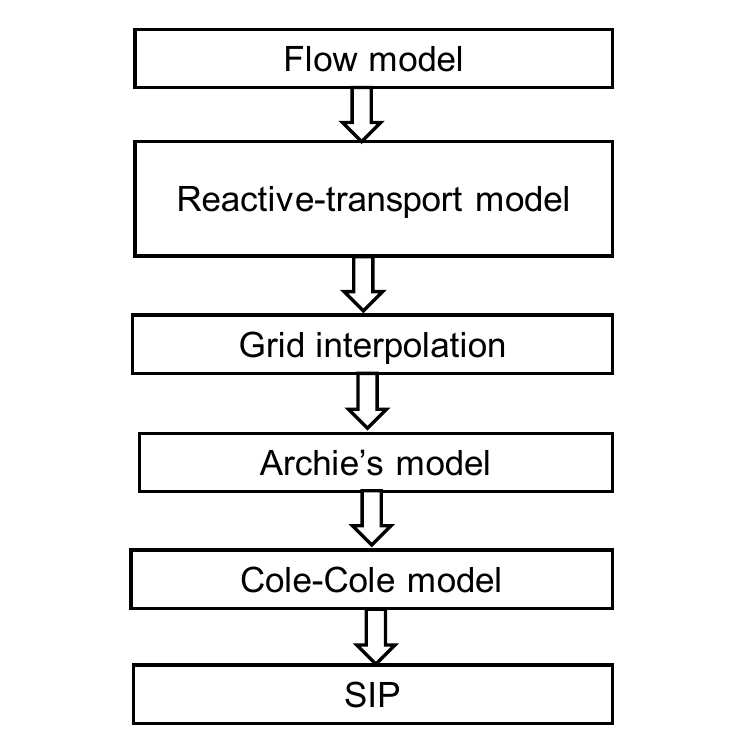}
  \caption{\textbf{\textrm{Coupling PFLOTRAN and SIP Process Models:}}~Steps involved in coupling
fluid flow, solute transport, and SIP process models in the PFLOTRAN-SIP framework;
further details are available in \cite{johnson20173,lichtner2015pflotran}. 
Details of the inputs, data, process models, and outputs are described in Algorithm~1.}
  \label{fig:flowchart}
\end{figure}

\texttt{PFLOTRAN} and \texttt{E4D} use Message Passing Interface calls for inter-process communication.
Based on user specification, \texttt{PFLOTRAN} divides the computing resources between \texttt{PFLOTRAN} and \texttt{E4D} at the initial step.
\texttt{PFLOTRAN} and \texttt{E4D} read their corresponding input files and complete pre-simulation steps.
These include setup of the flow model, the solute transport model, the SIP model, and the mesh interpolation matrix.
Mesh interpolation is needed for two reasons: (1)~the meshes of \texttt{PFLOTRAN} and \texttt{E4D} are different and (2)~the solution procedure of \texttt{PFLOTRAN} is based on the FVM while \texttt{E4D}'s solution procedure is based on the FEM.
As a result, the state variables (e.g., solute concentration, fluid saturation) computed at the cell center by \texttt{PFLOTRAN} need to be accurately transferred from the \texttt{PFLOTRAN} mesh to the \texttt{E4D} mesh to calculate electrical conductivities.
Generation of the mesh interpolation matrix is described in Sec.~\ref{minterp}.
Algorithm \ref{alg} and Fig.~\ref{fig:flowchart} summarize the coupling of \texttt{PFLOTRAN} and \texttt{SIP} models.

\subsection{Petrophysical Transformation}
\label{SubSec:S2_Petro_Transform}
To simulate SIP signals during fluid flow and solute transport, a mathematical relationship linking fluid flow state variables and bulk electrical conductivities is required.
Archie's law \cite{archie1942,glover2010generalized,shah2005generalized} is a petrophysical transformation relating state variables simulated by \texttt{PFLOTRAN} to bulk electrical conductivities:
\begin{equation}
\label{eq:Archies_Law}
  \sigma_\mathrm{b}\left(\mathbf{x}\right) = \frac{1}{\tau_\mathrm{f}} \boldsymbol{\mathrm{\phi}}^{\alpha} s_\mathrm{f}^{\beta} \sigma_\mathrm{f},
\end{equation}
where $\tau_\mathrm{f}$ [--] is the tortuosity factor (path length of current), $\sigma_\mathrm{b}(\mathbf{x})$ [S/m] is the bulk electrical conductivity at near-zero frequency ($\omega \sim 0$), $\alpha$ [--] is the cementation exponent (1.8 to 2.0 for sandstone), $s_\mathrm{f}$ [--] is the solute concentration simulated by \texttt{PFLOTRAN}, $\beta$ [--] is the saturation exponent (close to 2.0), and $\sigma_{\mathrm{f}}$ [S/m] is the fluid electrical conductivity.

To account for frequency dependence, Eq.~(\ref{eq:Archies_Law}) was modified using the Cole-Cole model \cite{colecole1941,cole1942dispersion,dias2000developments,tarasov2013use,revil2014spectral}:
\begin{equation}
\label{eq:Cole-Cole}
  \sigma^*\left(\mathbf{x},\mathrm{\omega}\right) = \sigma_b\left(\mathbf{x}\right) \left\{1 + \eta_\mathrm{a} \left[\frac{\left(i \omega t_\mathrm{r} \right)^{\gamma}}{1 + \left(1 - \eta_\mathrm{a} \right) \left(i \omega t_\mathrm{r} \right)^{\gamma} }\right]\right\},
\end{equation}
where $i^2 = -1$, $\gamma$ [--] is a shape parameter, and $t_\mathrm{r}$ [s] is the characteristic relaxation time constant (time for the imaginary electrical component to reach equilibrium after perturbation) related to characteristic pore or grain size.

\subsection{Mesh Interpolation}
\label{minterp}
Once the frequency-dependent real and imaginary components of bulk electrical conductivities were calculated on the \texttt{PFLOTRAN} mesh, they were interpolated onto the \texttt{E4D} mesh.
The conductivity at any intermediate point in a \texttt{PFLOTRAN} mesh cell was approximated using tri-linear interpolation.
Tri-linear interpolation is a multivariate interpolation function on a 3-dimensional regular grid.
It linearly approximates the value of a function at an intermediate point $(x, y, z)$ within the local rectangular prism, using function data on the lattice points.
Here, approximated values were computed using values at the \texttt{PFLOTRAN} cell centers surrounding the point at \texttt{E4D} grid~\cite{JOHNSON2017}.

\section{NUMERICAL MODEL SETUP}
\label{Sec:S3_model_framework}
\subsection{PFLOTRAN Model Setup}
\label{SubSec:PFLOTRAN_model_setup}
\begin{figure}[htbp]
  \centering
  \includegraphics[width=1.0\textwidth]{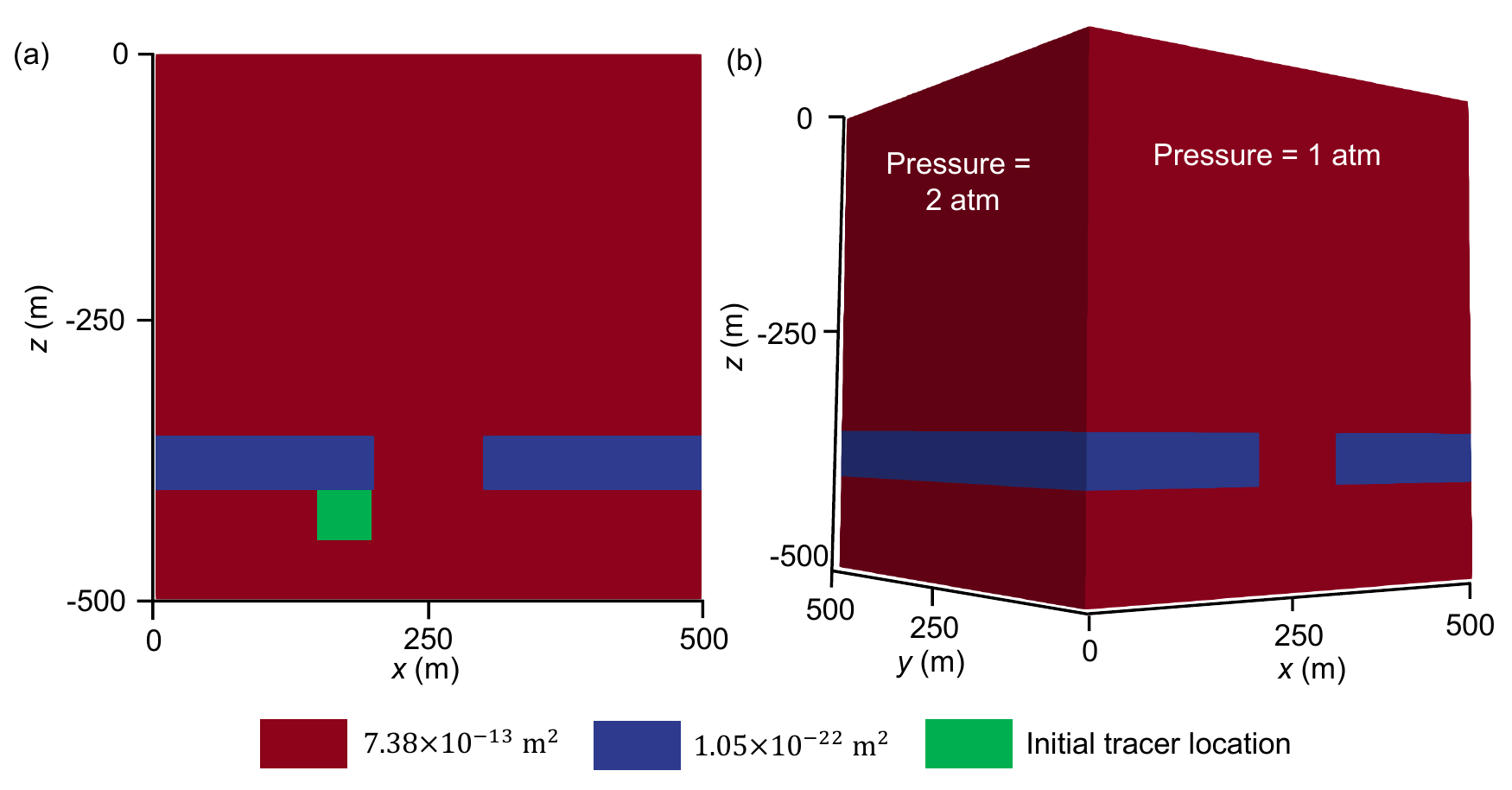}
  \caption{\textbf{PFLOTRAN model domain:}~Schematics of (a)~permeability distribution and (b)~pressure boundary conditions.}
  \label{fig:model_domain}
\end{figure}

A~simple example model was developed to demonstrate \texttt{PFLOTRAN-SIP}.
Similar to the Hanford Site, Richland, Washington \cite{JOHNSON2017}, a uniform pressure gradient drove flow in the positive $x$ direction.
The system was intended to be representative of sandstone with an intermittent shale layer.
This synthetic problem included contaminant transport with the intention to support remediation by providing insight into the evolution of the tracer distribution.
The domain was $500 \times 500 \times 500$\,m\textsuperscript3 and consisted of three layers as shown in Fig.~\ref{fig:model_domain}.
The \texttt{PFLOTRAN} mesh had a total of 125,000 finite volume cells.
The upper layer was $500 \times 500 \times 350$\,m\textsuperscript{3} and extended from $z$ = 0 to $-350$\,m as a highly conductive material with $\kappa=7.38\times10^{-13}$\,m\textsuperscript{2}.
The fluid was water and rock properties (e.g.,~$\kappa$, $\phi$, $D$, etc.) are representative of sandstone.
The middle layer was less permeable ($\kappa=1.05\times10^{-22}$\,m\textsuperscript2) with size $500 \times 500 \times 50$\,m\textsuperscript{3} extending from $z$ = $-350$ to $-400$\,m.
This $\kappa$ is representative of shale or granite.
The low-permeability layer, however, included a small-volume, sandstone ($\kappa=7.38\times10^{-13}$\,m\textsuperscript{2}) material between $x = 300$ and $350$\,m, $y = 0$ and $500$\,m, and $z = -400$ and $-450$\,m.
The bottom layer was also sandstone ($\kappa=7.38\times10^{-13}$\,m\textsuperscript{2}) with dimensions of $500\times500\times100$\,m\textsuperscript{3}.

A solute (conservative tracer) at $10 mol/\mathrm{kg}$ was placed below the low permeable zone as shown in Fig.~\ref{fig:model_domain} as the green $50 \times 500 \times 50$\,m\textsuperscript3 block.
The initial and boundary conditions for the model included:~pressure of 1\,atm at the top with a hydrostatic pressure gradient from top to bottom.
The left face ($x = 0$) was assigned a hydrostatic pressure of 2\,atm to drive flow from left to right.
For solute transport, the boundary conditions were zero-concentration Dirichlet inflow at the left face and zero diffusive gradient outflow at the right face that allowed only advective outflow.
The remaining faces were specified as zero-solute flux boundaries.

For low- and high-$\kappa$ zones, $\tau = 1$ while $\phi$ were 0.3 and 0.25, respectively.
Solute diffusivity was 10\textsuperscript{-9}\,{m$^2$/s}.
The Newton solver (20-iteration maximum) was applied for flow and solute transport.
For the flow solver, relative and absolute tolerances [--] were $10^{-50}$ with a relative update tolerance of $10^{-60}$ while for solute transport solver, relative and absolute tolerances were $10^{-4}$ with a relative update tolerance of $10^{-60}$.
The simulation was run for one year with an initial time step of $10^{-8}$\,years, which was allowed to accelerate by a factor of 8.

\subsection{SIP Model Setup}
\label{SubSec:SIP_model_setup}
Although the domain dimensions for SIP simulations were identical to the \texttt{PFLOTRAN} simulation, there was only a single layer.
The corresponding \texttt{E4D} mesh for the simulation had 86,780 nodes and 609,562 tetrahedral elements.
To avoid zero potentials effects on the SIP model, zero potentials were enforced on the external boundaries, which were 9,500\,m away from each lateral boundary  except for the top, which corresponded to the ground surface in both models.
This extension of the SIP model domain aided the SIP simulation \cite{johnsonetal2010}.
A~total of 80 point electrodes were placed in the domain, all located at $z= -425$\,m arranged in 5 lines along the $x$-axis, with each line comprising 16 electrodes.
The electrode coordinates started at $(40, 50, -425)$ and ended at $(460, 450, -425)$ with a 100\,m separations between lines see, Fig.~\ref{fig:sip_model_domain}.
Although in practice it is much easier to place electrodes on the surface, in this simulation they were placed in the region of interest (i.e.,~at depth) to provide more accurate data that facilitated a better inversion of subsurface properties and processes.
Compared to surface-lain electrodes, electrodes buried at depth are less impacted by noise (e.g.,~due to anthropogenic activities).
Electrode measurement configurations included a combination of Wenner and dipole-dipole arrays.

A~current of 1A was injected and received at a pair of electrodes, and the potential difference was measured at another pair of electrodes. 
There are various advantages of injecting and receiving the current through a pair of electrodes.
For example, such a measurement system can eliminate any inaccuracies caused by the injecting circuit impedance (the contact impedance between the probe and the medium, which can be high).
Using the prescribed measurement configuration, a total of 1,062 simulated measurements were collected to capture electrical impedance and phase shift.

The electrical conductivity of the fluid at $\omega = 0 \mathrm{Hz}$ was $2 \times 10^{-3}$\,S/m.
Parameters $\alpha$, $\beta$, and $t_\mathrm{r}$ were 0.564, 0.576, and 0.061\,s, respectively, all representative of sandstone \cite{titov2002theoretical}.
SIP analysis was performed for five different frequencies: 0.1, 1, 10, 100, and 1,000\,Hz.
Forward model simulations were performed using 61 processors, where 20 processors were assigned for \texttt{PFLOTRAN} and 41 for \texttt{E4D}.
Out of those 41 processors, 40 performed SIP simulations for different measurement configurations, and the remaining processor gathered the simulated data.

\subsection{SIP Inversion of electrical conductivity}
\label{SubSec:SIP_inversion}
For verification, \texttt{E4D}'s inversion module was used to estimate frequency-dependent electrical conductivity based on the simulated electrical impedance and phase-shift data.
This estimated conductivity was compared with the simulated conductivity generated by the \texttt{PFLOTRAN-SIP} framework.
The employed inversion process was blind (i.e., we did not provide prior constraints on the conductivity).
This can be improved by providing detailed conductivity information to \texttt{E4D}'s inversion module.
The SIP inversion employs an unstructured mesh, which consisted of 51,124 nodes with 316,183 mesh elements.
Low-order mesh elements were generated to make the inversion process simple and computationally efficient because high-order mesh elements did not improve the outcome \cite{e4dmanual}.
However, the meshes were refined around electrodes where the volume of each mesh was in the order of cm$^3$.
Simulated measurements (electrical impedance) by \texttt{PFLOTRAN-SIP} were the data supplied to the inversion process as observations.

\texttt{E4D} was inverted by minimizing the following objective function:
\begin{equation}
  \label{eq:inversion_objective}
  \Phi = \Phi_\mathrm{d} \left[\mathbf{W}_\mathrm{d}\mathrm{\left(\Phi_{obs} - \Phi_{pred}\right)} \right] + \zeta \Phi_\mathrm{m} \left[\mathbf{W_\mathrm{m}}\left(\boldsymbol{\sigma}_\mathrm{est} - \boldsymbol{\sigma}_\mathrm{ref}\right) \right],
\end{equation}
where $\Phi_\mathrm{d}$ is a scalar operator that quantifies the misfit between observed and simulated data (e.g.,~electrical impedance and phase shift) based on the user-specified norm (e.g.,~Euclidean norm), $\Phi_\mathrm{m}$ is another operator that provides a scalar measure of the difference between the frequency-dependent electrical conductivity distribution, $\boldsymbol{\sigma}_\mathrm{est}$\,[S/m], and constraints placed upon the structure of $\sigma_{\mathrm{ref}}$\,[S/m], $\zeta$ is the regularization parameter, $\mathbf{W}_{\mathrm{d}}$ is the data-weighting matrix, and $\mathbf{W}_{\mathrm{m}}$ is the model-weighting matrix.
$\sigma_{\mathrm{est}}$ and $\sigma_{\mathrm{ref}}$ are the estimated and reference frequency-dependent electrical conductivities.
The user specified bounds on the frequency-dependent conductivity in each mesh cell were 0.00001 and 1.0.
The $\Phi_{\mathrm{obs}}$ and $\Phi_{\mathrm{pred}}$ were the observed and simulated data, respectively.
Eq.~(\ref{eq:inversion_objective}) is solved using the iteratively reweighted least square method \cite{scales1988robust}.
Further details on the parallel inverse modeling algorithm and its implementation in \texttt{E4D} are available \cite{johnsonetal2010}.

The $\zeta$ value was 100 at the beginning of the inversion and decreased as the nonlinear iteration progressed.
Before $\zeta$ was reduced, the minimum fractional decrease in the objective function, $\Phi$, between iterations had to be less than 0.25 upon which $\zeta$ was reduced to 0.5.
The convergence of the SIP inversion procedure was based on the $\chi^2$ value of the current iteration after data culling, computed as:
\begin{equation}
  \label{eq:inv_obj_function}
  \chi^2 = \frac{\Phi_\mathrm{d}}{n_\mathrm{d} - n_{\mathrm{c}}},
\end{equation}
where the data residual is the difference between observed and estimated values divided by the standard deviation for that measurement.
$n_\mathrm{d}$ is the total number of survey measurements and $n_\mathrm{c}$ is the number of measurements selected from the total number of measurements during the current iteration.


\section{RESULTS \& DISCUSSION}
\label{Sec:S4_Results}
\begin{figure}[htbp]
  \centering
  \includegraphics[width=0.5\textwidth]{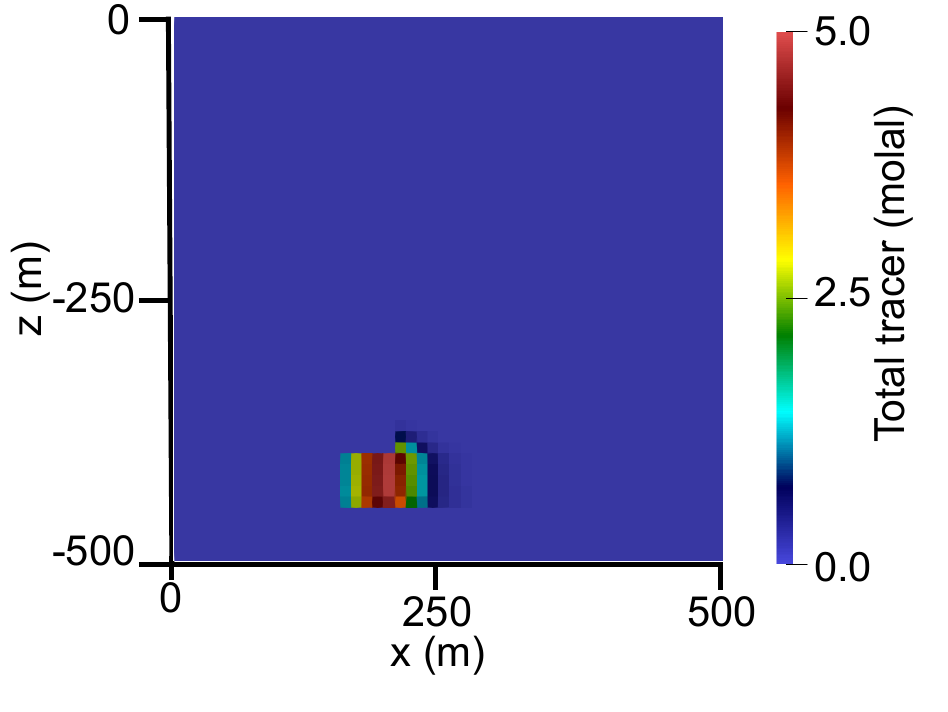}
  \caption{\textrm{\textbf{PFLOTRAN Simulation:}} Spatial distribution of tracer concentrations after one year.}
  \label{fig:tracer_conc}
\end{figure}

The one-year \texttt{PFLOTRAN-SIP} model simulations were completed in two minutes. 
The computation was performed on 61 Intel\textsuperscript{\textregistered}~Xeon\textsuperscript{\textregistered} CPU~E5-2695~V4\,@~2.1GHz processors.
Fig.~\ref{fig:tracer_conc} shows the tracer concentrations at the end of simulation.
In one year, the pressure gradient drove the tracer about 100\,m from its initial location in the $x$-direction and also moved it upward about 20\,m.

\begin{figure}[htbp]
  \centering
  \includegraphics[width=1.0\textwidth]{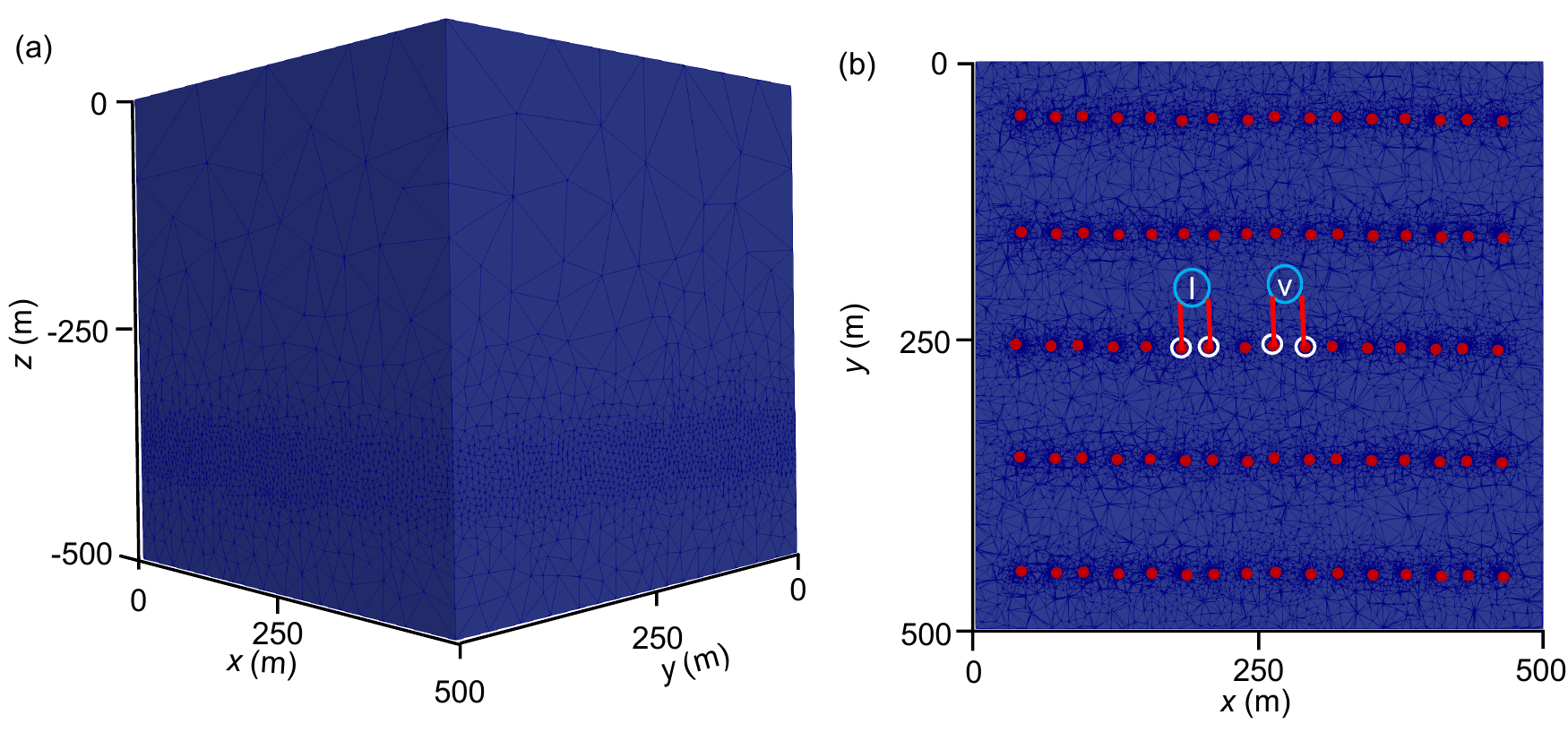}
  \caption{\textbf{SIP model:}~(a)~The 3D SIP model domain where (b)~red dots represent electrodes on the $xy$-plane. White circles represent electrode configuration of 80$^\mathrm{th}$ out of 1,062 simulated electrical impedance where I and V represent current and potential electrodes, respectively.}
  \label{fig:sip_model_domain}
\end{figure}

\begin{figure}[htbp]
  \centering
  \includegraphics[width=1.0\textwidth]{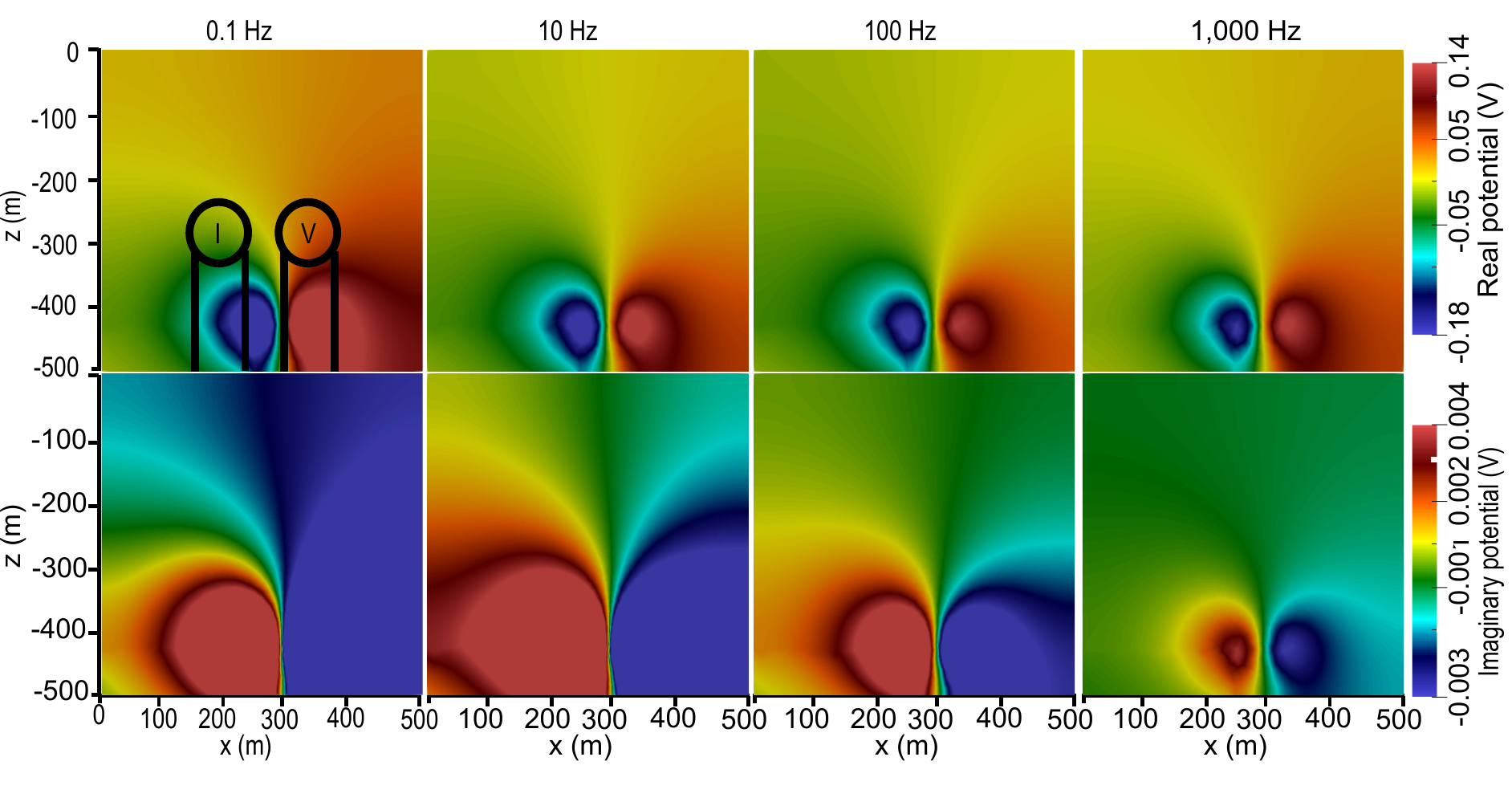}
  \caption{Slices of simulated real (top) and imaginary (bottom) components of complex electrical potentials/impedances at $y = 250$\,m for a single measurement after one year. Measurement location is in top left-corner plot of this figure and in Fig.~\ref{fig:sip_model_domain}(b).}
  \label{fig:potential}
\end{figure}

The SIP module in \texttt{PFLOTRAN-E4D} simulated real and imaginary electrical impedance at 0.1, 1, 10, 100, and 1,000\,Hz.
Because of minimal differences between 1 and 10\,Hz, only results for 0.1, 10, 100, and 1,000\,Hz are discussed.
This indicated that some frequencies may be redundant because they yield similar impedance.
Sensitivity analyses can be performed to identify redundant frequencies; however, this was beyond the scope of this paper.
Fig.~\ref{fig:potential} shows the real and imaginary potentials due to changes in tracer concentration for the various frequencies. 
Also, this figure provides information on the change in electrical potential at different frequencies for a single measurement configuration, indicating the maximum tracer concentration.
The 80\textsuperscript{th} out of 1,062 electrical impedance measurements (see, Fig.~\ref{fig:sip_model_domain}(b)) was selected where tracer concentrations were most evident.
The response clearly shows the polarization feature of the tracer. 
The gradient of the real electrical potential was high near $x = 300$\,m (top row of Fig.~\ref{fig:potential}) where tracer concentrations were maximum. 
From Fig.~\ref{fig:potential}, it is evident that the real potential response for 0.1\,Hz was different from the responses at 10, 100, and 1,000\,Hz. 
The root-mean-square error (RMSE) between these responses was approximately 15\% of the maximum real potential value indicating that frequency has an impact on the real potential distribution.

The bottom row of Fig.~\ref{fig:potential} shows imaginary component of complex electrical potential responses where the polarity was switched (colors interchanged).
Unlike the real electrical potential, each imaginary electrical potential was notably different indicating its frequency dependence.
The corresponding RMSE between responses was $\sim$85\% of the maximum imaginary potential value.
Such high variation was expected as the imaginary electrical potential depends on frequency, chargeability, and relaxation time, although the last two were constant in this study.
Because the response of the imaginary potential was clearly visible in the simulation, this indicated that the \texttt{PFLOTRAN-SIP} framework can effectively simulate polarized geologic materials.

\begin{figure}[htbp]
  \centering
  \includegraphics[width=1.0\textwidth]{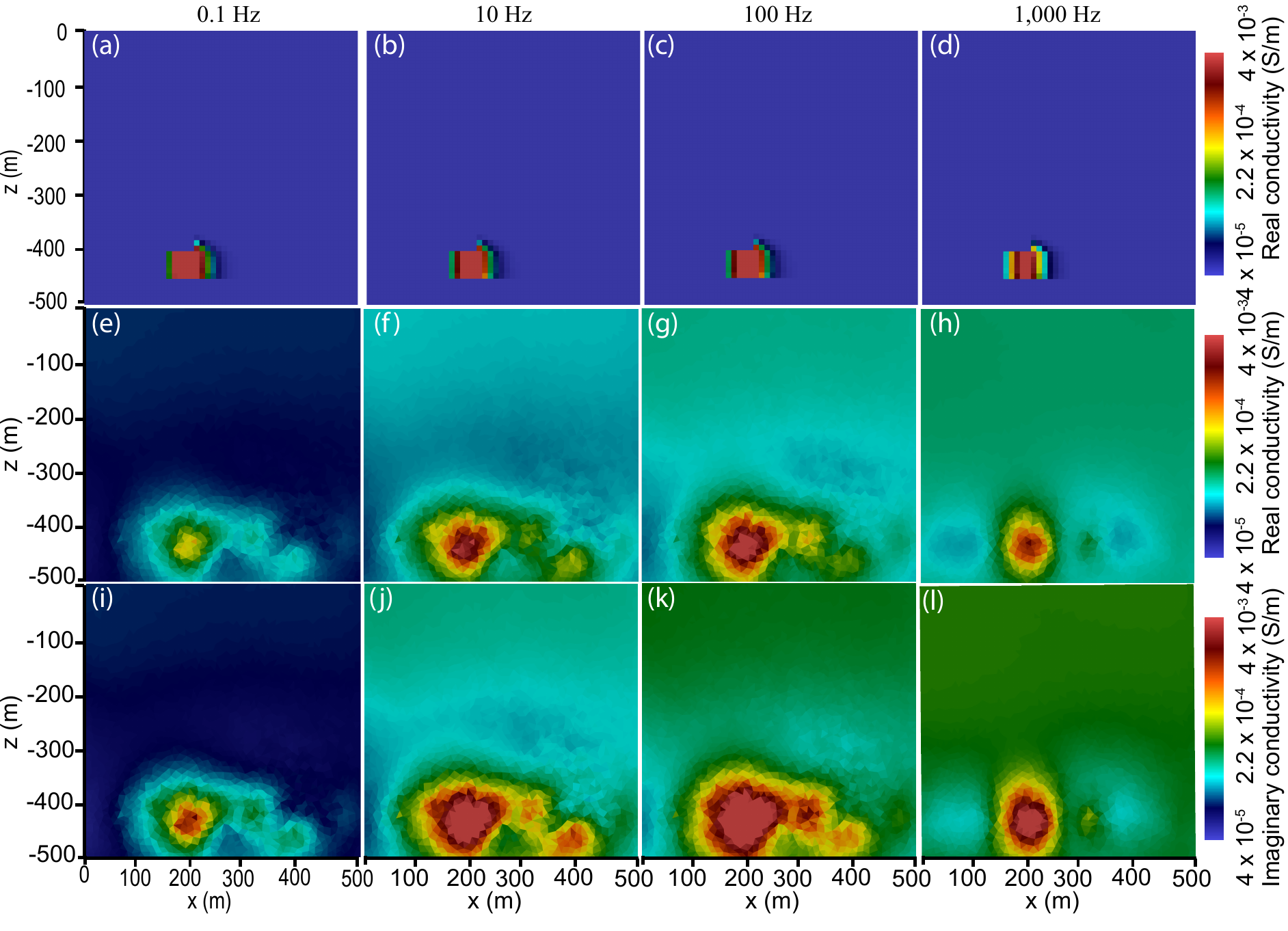}
  \caption{Simulated and estimated frequency-dependent electrical conductivities at $y = 250$\,m after one year. (a)--(d)~True-electrical conductivities from the \texttt{PFLOTRAN-SIP} framework, (e)--(h) estimated bulk-real conductivities from SIP inversion, and (i)--(l)~imaginary components of estimated bulk complex electrical conductivities from SIP inversion.}
  \label{fig:real_and_complex_conductivity}
\end{figure}

Fig. \ref{fig:real_and_complex_conductivity} shows the simulated and estimated frequency-dependent electrical conductivities using the \texttt{PFLOTRAN-SIP} framework with the SIP inversion module in \texttt{E4D}. 
The true (\texttt{PFLOTRAN} simulated) and estimated (inversion of survey data) real electrical conductivities are plotted in Fig.~\ref{fig:real_and_complex_conductivity}(a)--(d) and Fig.~\ref{fig:real_and_complex_conductivity}(e)--(h), respectively.
SIP inversion was performed using the simulated electrical impedance, and phase-shift data obtained from \texttt{PFLOTRAN-SIP} model runs after one year. Inversion converged after 48 iterations when $\chi^2$ reached 60.
The computational time required to perform SIP inversion was approximately two hours on 41 Intel\textsuperscript{\textregistered}~Xeon\textsuperscript{\textregistered}~CPU~E5-2695~V4 processors running at 2.1\,GHz.
Estimated electrical conductivities showed high contrast around the high tracer distribution/simulated conductivities, although they were more diffuse than the true (simulated) distribution (Fig.~\ref{fig:real_and_complex_conductivity}(a)--(h)).
ERT provided data for Fig.~\ref{fig:real_and_complex_conductivity}(e), but SIP provided data for Fig.~\ref{fig:real_and_complex_conductivity}(e)--(l). Although not all SIP data were informative, some were useful. 
For example, estimated conductivities at 1,000\,Hz were more accurate than frequencies \textless1,000\,Hz with the same inversion constraints.
Later, real conductivity values were used in Eq.~(\ref{eq:Cole-Cole}) to provide initial guesses for imaginary conductivities for SIP inversion.
Estimated imaginary electrical conductivity distributions are shown in Fig.~\ref{fig:real_and_complex_conductivity}(i)--(l).
Similar to estimated real conductivities, imaginary conductivities computed from SIP inversion were also diffuse.
The inversion process could be improved by providing prior information and structural constraints on electrical conductivities.
However, both estimated conductivities were generally consistent with the tracer distribution, which showed that the SIP inversion module can simulate electrical impedance and phase-shift data.
To summarize, SIP provides a major benefit over ERT in the form of greater information content.
This is because an SIP survey yields multiple datasets at different frequencies that help to overcome false positives i.e.,~indication of a tracer where none is present).
For example, from Fig.~\ref{fig:real_and_complex_conductivity} it is evident that the SIP inversion analyses at different frequencies revealed the same tracer region (not a false positive).
With an ERT survey, it may be difficult to identify a false positive from a true positive because ERT only generates a single dataset.

\begin{figure}[htbp]
  \centering
  \includegraphics[width=1.0\textwidth]{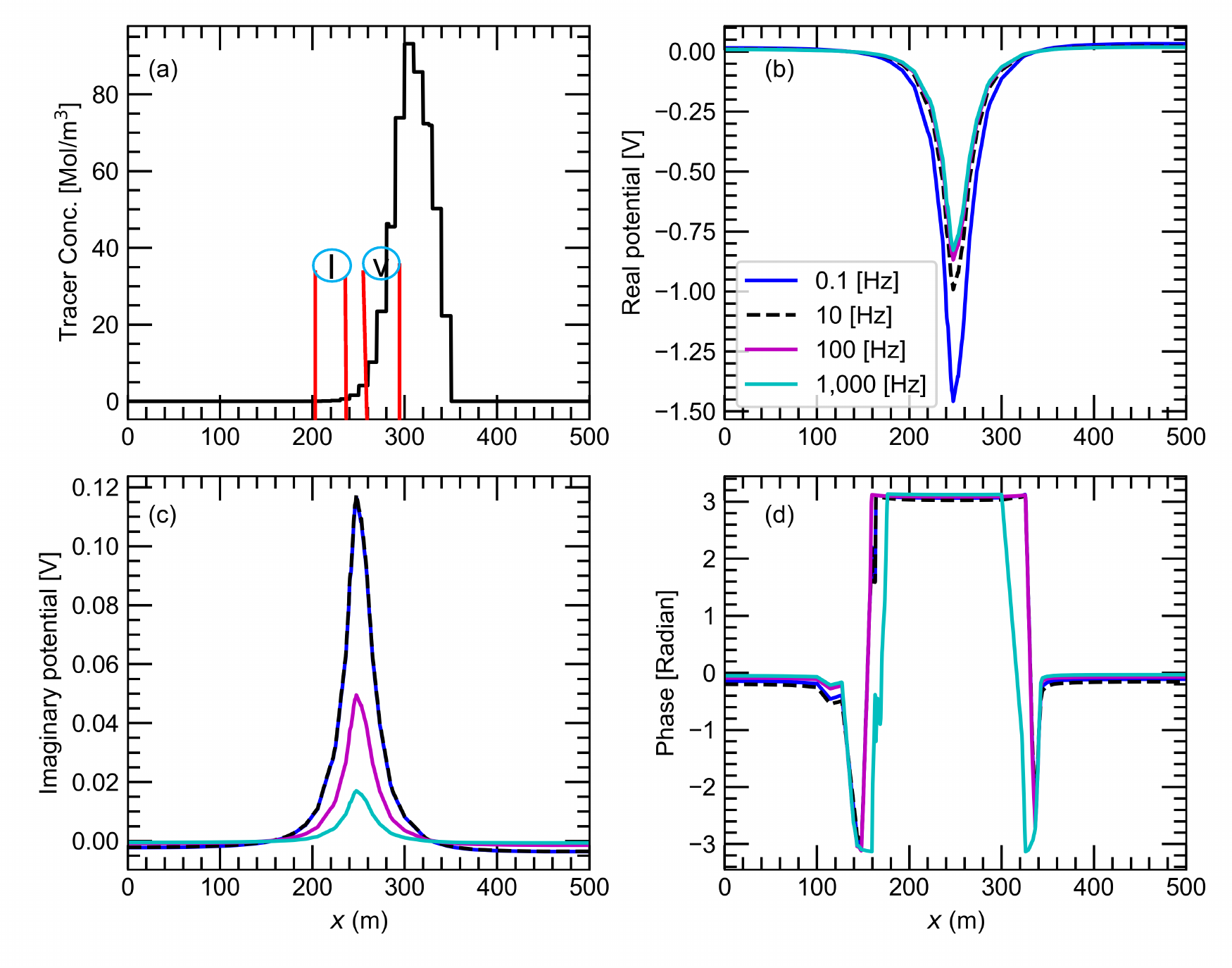}
  \caption{Distribution of (a)~tracer concentration and (b)~real potential, (c)~imaginary component of complex potential, and (d)~phase shift along the $x$-axis at $y = 250$\,m and $z = -425$\,m.}
  \label{fig:tracer_poten_phase}
\end{figure}

Fig.~\ref{fig:tracer_poten_phase}(a)--(c) show simulated outputs of tracer concentrations, real potentials, and imaginary potentials for the 80-electrode measurement configuration at frequencies of 0.1, 10, 100, and 1,000\,Hz.
The location of maximum tracer concentration was around $x = 300$\,m (Fig.~\ref{fig:tracer_poten_phase}(a)).
The locations of current and potential measurement electrodes were at ($x = 208,\,236,\,264,$ and $292$\,m, $y = 250$\,m, and $z = -420$\,m) (Fig.~\ref{fig:sip_model_domain}(b)).
Note that electrodes were not placed at the location of maximum concentration, but they were placed 50\,m right of maximum concentration in a line in the subsurface.
Nevertheless, the measured potentials provided meaningful information on the bounds of the tracer distribution as well as revealing the significance of higher frequencies obtained from a combination of electrical impedance and phase shift.

\begin{figure}[htbp]
  \centering
  \includegraphics[width=1.0\textwidth]{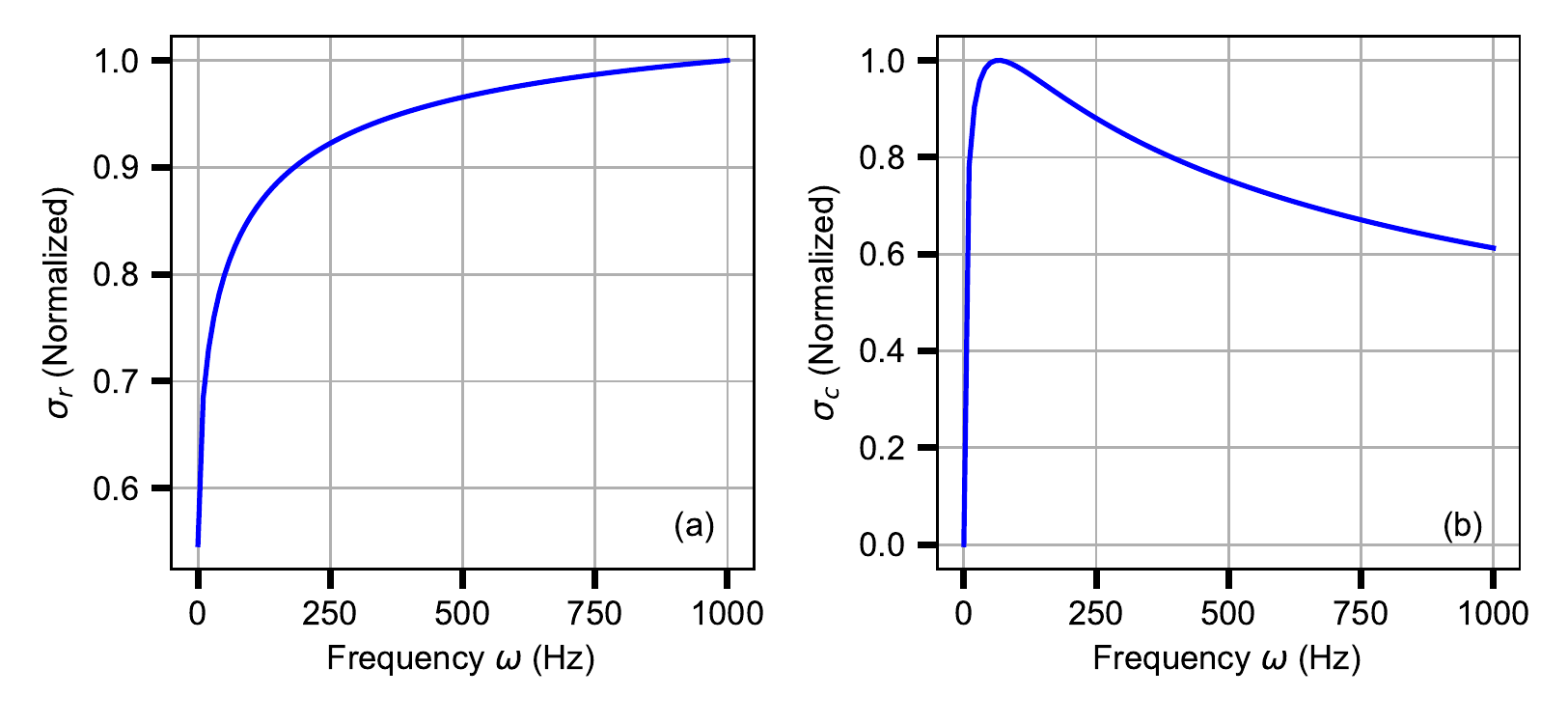}
  \caption{(a)~Real, $\sigma_r$, and (b)~imaginary, $\sigma_c$, components of complex conductivities vs. frequency.
  Each component of the complex conductivity was normalized with respect to its maximum value to better show trends.}
  \label{fig:FreqvsConds}
\end{figure}

For this study, only tracer concentration impacted real and imaginary components of complex conductivities revealed through the Cole-Cole model because $\alpha$, $\beta$, and $t_\mathrm{r}$ were held constant to investigate the effect of tracer concentration over different frequencies. Fig.~\ref{fig:FreqvsConds}(a) and (b) show how the Cole-Cole model increased real conductivities and decreased the imaginary component of complex conductivities over different frequencies.
Fig.~\ref{fig:tracer_poten_phase}\,(b)~and~(c) shows that the absolute real potential and imaginary potential decreased as frequency increased.
As noted in Sec.~\ref{SubSec:S2_E4D_SIP_Model} and in by \cite{johnson20173}, for SIP simulations, \texttt{E4D} first solved the real potential, $\Phi_\mathrm{r}$.
That is, $-\mathrm{div}\left[\sigma_\mathrm{r} \mathrm{grad}\left[\Phi_\mathrm{r}\right] \right] = I$, where $\sigma_\mathrm{r}$ is the real component of $\sigma^*(\mathbf{x},{\omega})$ and $\Phi_\mathrm{r}$ is inversely proportional to $\sigma_\mathrm{r}$.
Also, $\sigma_\mathrm{r}$ increased as $\omega$ increased; hence the absolute value of the real potential distribution (as shown in Fig.~\ref{fig:tracer_poten_phase}\,(b)) decreased as $\omega$ increased.
After $\sigma_\mathrm{r}$ was evaluated, \texttt{E4D} computed the complex potential by solving $\mathrm{div}\left[\sigma_\mathrm{r} \mathrm{grad}\left[\Phi_\mathrm{c}\right] \right] = -\mathrm{div}\left[\sigma_\mathrm{c} \mathrm{grad}\left[\Phi_\mathrm{r}\right] \right]$ where $\sigma_\mathrm{c}$ is the imaginary part of $\sigma^*(\mathbf{x},{\omega})$ and $\Phi_\mathrm{c}$ is the imaginary potential.
Thus, $\Phi_\mathrm{c}$ is proportional to $\sigma_\mathrm{c}$.
Also, $\sigma_\mathrm{c}$ decreased as $\omega$ increased; hence the absolute value of the imaginary potential distribution (as shown in Fig.~\ref{fig:tracer_poten_phase}\,(c)) decreased as $\omega$ increased.
  
Fig.~\ref{fig:tracer_poten_phase}\,(d) shows phase-shift data distribution along the same line as the tracer distribution, real and imaginary potential distribution.
Mathematically, the phase shift is the inverse tangent of the ratio between imaginary and real potential responses.
Physically, it is the shift between measured voltage and applied current signals that is largely governed by the polarization characteristics of the subsurface.
In this study, phase shift leveraged signals from both real and imaginary potential responses to improve the interpretation of complex electrical impedance.
From Fig.~\ref{fig:tracer_poten_phase}, there was a change in phase shift where tracer transport was predominant.
Moreover, the 1,000\,Hz frequency bounded the tracer zone better than lower frequencies that cannot be distinguished with ERT. 
This phase shift helped constrain the polarized region or bound the interface between tracer-laden and tracer-free fluids. 
After identifying the region of interest according to these constraints, further geoelectrical interrogation could be performed with this volume.
It is critical to mention that accurate estimation of the region of interest was found without performing a computationally expensive numerical inversion.
Hence, through phase-shift signatures across multiple frequencies, the \texttt{PFLOTRAN-SIP} framework facilitates the identification of polarized or geochemically altered zones. 

IP arises from solute transport and accumulation of ions/electrons in polarized materials \mbox{(e.g.,~those} with different grain types, colloids, biological materials, phase-separated polymers, blends, and crystalline minerals, etc.) when subject to an external electric field. 
Five mechanisms govern IP phenomena at frequencies $< 1$\,MHz: (1)~Maxwell-Wagner polarization, which occurs at high frequencies \cite{alvarez1973complex,chelidze1999electrical,lesmes2001dielectric,chen2006geometrical}; (2)~polarization of the inner part of the interface between minerals and water \cite{de1992generalized,leroy2009mechanistic,vaudelet2011induced,revil2012spectral}; (3)~polarization of the outer part of the interface between minerals and water \cite{dukhin1974dielectric,de1992generalized}; (4)~membrane polarization for multi-phase systems \cite{marshall1959induced,vinegar1984induced,titov2002theoretical}; and (5)~electrode polarization observed in the presence of disseminated conductive minerals such as sulfide minerals and pyrite \cite{wong1979electrochemical,merriam2007induced,seigel2007early}.

Our \texttt{PFLOTRAN-SIP} simulations were geared toward IP mechanisms (1), (4), and (5).
To simulate mechanisms (2) and (3), Eq.~(\ref{eq:Cole-Cole}) must be replaced with conductivity models that account for interface polarization with consideration of effective pore size, electrical formation factor, distribution of relaxation times, and sorption mechanisms \cite{revilandflorsch2010,vaulet2011}.
Note that the Cole-Cole model given by Eq.~(\ref{eq:Cole-Cole}) neglects the effects of polarization at interfaces or sorption onto mineral surfaces.
The \texttt{PFLOTRAN-SIP} framework can easily account for such modifications in conductivity, but this is a future endeavor.

\section{CONCLUSIONS}
\label{Sec:S6_Conclusions}
This work demonstrated the \texttt{PFLOTRAN-SIP} framework, which simultaneously simulates fluid flow, reactive transport, and SIP.
A reservoir-scale tracer transport model demonstrated the proposed \texttt{PFLOTRAN-SIP} framework where fluid flow and tracer concentration evolution were simulated over one year.
Then, we simulated 1,062 electrical impedance at four frequencies.
These simulations showed that contrast in real potential were minimal even as the frequency varied.
However, there was a significant change in contrast of complex potentials across frequencies.
Phase shift (a combination of real and complex potentials) helped identify the region where tracer concentration was high.
This analysis showed that SIP has two major advantages over ERT.
First, SIP provides frequency-dependent electrical impedance data.
Second, phase-shift signatures obtained from SIP analysis identified and constrained geochemically altered zones.
Combining frequency-dependent real potential, complex potential, and phase responses from an SIP survey/simulation paints a more detailed picture of the subsurface with an enhanced ability to detect contaminants/tracers. 
Moreover, coupling fluid flow, reactive transport, and SIP models can better detect contaminants compared to either the ERT or SIP method alone.
For instance, through our numerical example, solute transport simulations provided insight into the tracer distribution.
This information was used to customize SIP inversion to estimate frequency-dependent electrical conductivities, which yielded an improved image of tracer concentrations at different frequencies.
Although this work focused on simulating tracer transport, it could also be applied to detect hydrocarbon flow, changes in the subsurface due to geochemical reactions, sulfide minerals, metallic objects, municipal wastes, and salinity intrusion.
Moreover, this code could be used in feasibility studies for developing waste sequestration sites.
 \section*{Acknowledgments}
This research was funded by the U.S. Department of Energy (DOE) Basic Energy Sciences (BES) and Fossil Energy (FE) programs.
MKM, SK, and BA also thank the support from Center for Space and Earth Sciences (CSES) Emerging Ideas R\&D Program.
The authors thank Glenn Hammond (Pacific Northwest National Laboratory) and Tim Johnson (Pacific Northwest National Laboratory) for the coupled framework \texttt{PFLOTRAN-E4D} upon which \texttt{PFLOTRAN-SIP} was built.
Los Alamos National Laboratory is operated by Triad National Security, LLC, for the National Nuclear Security Administration of U.S. Department of Energy (Contract No. 89233218CNA000001).
The authors also want to acknowledge the comments provided by the anonymous reviewers and editors that substantially improved the manuscript.

\section*{Conflict of Interest}
The authors declare that they do not have any conflicts of interest.
\section*{Computer Code Availability, Installation, and Contribution}
The \texttt{PFLOTRAN-SIP} source code is available for download at \url{https://bitbucket.org/satkarra/pflotran-e4d-sip/src/master/}.
\texttt{E4D} source code can be downloaded at \url{https://bitbucket.org/john775/e4d_dev/wiki/Home}.
The \texttt{PFLOTRAN-SIP} simulation input files used for this manuscript are available in the public github repository \url{https://github.com/bulbulahmmed/PFLOTRAN-SIP}.
Additional information regarding the simulation datasets can be obtained from Bulbul Ahmmed (Email:~\texttt{bulbul\_ahmmed@baylor.edu}) and Maruti Kumar Mudunuru (Email: \texttt{maruti@lanl.gov}).
\newline
\textbf{Installation of the code}
\begin{enumerate}
    \item Dowload PETSC: git clone \url{https://gitlab.com/petsc/petsc}
    \item cd petsc
    \item git checkout xsdk-0.2.0
    \item ./configure --CFLAGS=`-O3' --CXXFLAGS=`-O3' --FFLAGS=`-O3' --with-debugging=no --download-mpich=yes --download-hdf5=yes --download-fblaslapack=yes --download-metis=yes --download-parmetis=yes
    \item export PETSC\_DIR=/home/username/path\_to\_top\_level\_petsc
    \item export PETSC\_ARCH=gnu-c-debug
    \item cd \$PETSC\_DIR
    \item make all (or follow make instructions printed at the end of configuration.)
    \item Download PFLOTRAN: git clone \url{https://bitbucket.org/satkarra/pflotran-e4d-sip/src/master/}
    \item cd src/pflotran/
    \item make pflotran (for using multiple processors use \texttt{make -j np pflotran} (np = number of processors))
\end{enumerate}
\textbf{Code Contribution} \newline
\texttt{PFLOTRAN-SIP} was developed to simulate fully coupled flow, reactive-transport, and SIP processes.
\texttt{PFLOTRAN-SIP}  requires no external scripts so there is no impact on  computational performance.
The code, examples, and instructions for implementation are available at: \url{https://bitbucket.org/satkarra/pflotran-e4d-sip/commits/}.
\section{Figures and Algorithm}

\label{alg}
\begin{algorithm}
  \caption{{\small Overview of the proposed \texttt{PFLOTRAN-SIP} framework for simulating electrical impedance data}}
  \label{alg:coupled_framework}
  \begin{algorithmic}[1]
    \STATE{INPUT:~Initial and boundary conditions for fluid flow and solute transport models in \texttt{PFLOTRAN}, fluid density, porosity, saturation, volumetric source/sink with its location, intrinsic and relative permeabilities, dynamic viscosity, mass flow rate, diffusion/dispersion coefficients, tortuosity, solute source/sink with its location, Archie's and Cole-Cole model parameters, total simulation time, time-step for \texttt{PFLOTRAN}, interrogation frequencies, electrode locations and measurement configuration, number of processors for \texttt{PFLOTRAN} and \texttt{E4D}, and meshes for \texttt{PFLOTRAN} and \texttt{E4D}.}
    \STATE{Solve Eqs.~(\ref{eq:Richards_eqn})--(\ref{Eqn:SourceSink_Richards_Eqn}) for fluid pressure, fluid saturation, and fluid velocity.}
    \STATE{Solve Eq.~(\ref{eq:advective_dispersion}) to calculate the spatio-temporal distribution of solute concentration.}
    \STATE{Transfer solute concentration from \texttt{PFLOTRAN} to the \texttt{E4D} master processor to perform SIP simulations at specific times.}
    \STATE{Receive numerical model setup information from \texttt{PFLOTRAN} input files to perform mesh interpolation for SIP simulations.}
    \STATE{Broadcast run information and distribute mesh assignments to \texttt{E4D} slave processors.}
    \STATE{Calculate the mesh interpolation matrix to interpolate \texttt{PFLOTRAN} simulation outputs (e.g., solute concentrations) onto the \texttt{E4D} mesh for SIP simulations.}
    \STATE{Calculate electrical conductivities using Archie's model Eq.~(\ref{eq:Archies_Law}).}
    \STATE{Calculate frequency-dependent electrical conductivities using the Cole-Cole model Eq.~(\ref{eq:Cole-Cole}).}
    \STATE{Pass real and imaginary conductivities calculated at different frequencies to the \texttt{E4D} master processor to perform SIP simulations.}
    \STATE{Broadcast real and imaginary conductivities to \texttt{E4D} slave processors to compute pole solutions for electrode configurations.}
    \STATE{Solve Eq.~(\ref{eq:ip_potential_field_freq}) to compute complex electrical potential at different frequencies and solute concentrations at specified times.}
\end{algorithmic}
\end{algorithm}
%


\bibliographystyle{unsrt}
\bibliography{Master_References/References}

\end{document}